\documentclass[aps, notitlepage, reprint, showpacs, showkeys]{revtex4-1}
\usepackage[colorlinks=true,citecolor=blue,urlcolor=blue,linkcolor=blue]{hyperref}
\usepackage{amsmath,amssymb}
\usepackage{braket}
\usepackage{hyperref}
\usepackage[utf8]{inputenc}
\usepackage[]{graphicx}
\usepackage{tikz}
\usetikzlibrary{positioning,fit,arrows}
\usepackage{siunitx}
\usepackage{color}

\graphicspath{{./figures/}}
\newcommand{\EXP}[1]{\mathbb{E} \left[ #1 \right]}

\begin{document}

\title{Theory and experiments of disorder-induced resonance shifts and
  mode edge broadening in deliberately disordered photonic crystal
  waveguides} 
\date{\today} 
\author{Nishan Mann$^{1,*}$,Alisa
  Javadi$^2$, P.D. Garc\'{i}a$^2$, Peter Lodahl$^2$, and Stephen
  Hughes$^1$} \email{nmann@physics.queensu.ca}
\affiliation{$^1$Department of Physics, Queen's University, Kingston, 
  Ontario, Canada, K7L 3N6\\
  $^2$Niels Bohr Institute, University of Copenhagen, Blegdamsvej 17,
  DK-2100, Copenhagen, Denmark }

\begin{abstract}
  We study both theoretically and experimentally the effects of
  introducing deliberate disorder in a slow-light photonic crystal
  waveguide on the photon density of states. We first introduce a
  theoretical model that includes both deliberate disorder through
  statistically moving the hole centres in the photonic crystal
  lattice and intrinsic disorder caused by manufacturing
  imperfections. We demonstrate a disorder-induced mean blueshift and
  an overall broadening of the photonic density of states for various
  amounts of deliberate disorder. By comparing with measurements from
  a GaAs photonic crystal waveguide, we find good qualitative
  agreement between theory and experiment which highlights the
  importance of carefully including local field effects for modelling
  high-index contrast perturbations.  Our work also demonstrates the
  importance of using asymmetric dielectric polarizabilities for
  modelling positive and negative dielectric perturbations when
  modelling a perturbed dielectric interface in photonic crystal
  platforms.
\end{abstract}

\pacs{42.70.Qs, 42.25.Fx, 42.82.Et, 42.81.Dp}
\keywords{photonic crystal waveguides; scattering; disorder}
\maketitle

\section{\label{sec:introduction}Introduction}
% Intro to PC and applications
Photonic crystal (PC) cavities and waveguides are attractive
nanophotonic platforms for controlling and studying fundamental
light-matter interactions.  Aided by the presence of a photonic band
gap (PBG), which arises from the underlying periodic dielectric
structure, light within a PC cavity or waveguide is strongly confined
within a small volume or area. In the case of a PC waveguide (PCW),
light can be slowed down by orders of magnitude compared to a typical
slab or ridge waveguide, which increases the local density of photonic
states (LDOS). The ability to control light-matter interactions in PC
platforms leads to a host of photonic applications and rich optical
interactions \cite{Krauss2008, Lodahl2013}. For example, PC cavities
have been used for exploring cavity quantum electrodynamics
(cavity-QED) in both the weak and strong coupling regimes
\cite{Yoshie2004, Hennessy2007}, while PCWs have been exploited to
realize on-chip single photon sources
\cite{Lund-Hansen2008,Yao2009,Laucht2012}. Slow-light in PCWs also
enhance non-linear processes including pulse compression and soliton
propagation \cite{Colman2010}, third-harmonic generation
\cite{Monat2011} and four-wave mixing \cite{Li2012}. In addition, PCWs
have been integrated in various photonic circuits, as optical sensor
elements for refractive index measurements in biosensing
\cite{Skivesen2007} and chemical fluid detection
\cite{Topolancik2003}.

% Disorder and it's consequences.
In practice, PCWs are highly sensitive to manufacturing imperfections
(intrinsic disorder) which is inevitably introduced at the fabrication
stage. Disorder-induced losses are particularly detrimental in the
slow-light regime \cite{Baba2008}, which was predicted theoretically
by Hughes \emph{et al.} \cite{Hughes2005} using a photonic Green
function approach and is now a common finding of various similar
theoretical works in the literature \cite{Gerace2004, Wang2008,
  Song2010, OFaolain2007}. With continued improvements in
semiconductor fabrication techniques, and improved theoretical
understanding about how to mitigate disorder-induced losses
\cite{Wang2012,Mann2013}, various groups have experimentally
demonstrated PCW designs that have reduced disorder-induced losses
\cite{OFaolain2010, Sancho2012}.
%Recently, Sancho \emph{et.al.} demonstrated a PCW
%design with lower lossws with was explained via Bloch mode enginnering
%[my paper].

However disorder in PCs is not necessarily a hindrance as was noticed
by John in 1987 \cite{John1987}, who proposed disordered PCs for
experimentally observing the well known phenomena of Anderson
localization. Topolancik \emph{et al.} \cite{Topolancik2007}
experimentally demonstrated spectral peaks bearing signatures of
Anderson localization arising from localized modes in a deliberately
disordered PCW. Patterson \emph{et al.} \cite{Patterson2009b} utilized
coupled mode theory to highlight the effect of light localization via
multiple scattering by examining the transmission through a PCW in the
slow-light regime. The strong localized resonances in disordered PCWs
have also been used to enhance the spontaneous emission factor of
embedded quantum dots \cite{Sapienza2010}, and recently Thyrrestrup
\emph{et al.} \cite{Thyrrestrup2012} have proposed coupling quantum
dot emitters to a disordered PCW as a promising platform for
conducting QED experiments. Other applications of disordered PCWs
include enhanced light harvesting and random lasing \cite{Liu2014}.

%-----Theoretical Study of resonance shifts and modes------------------%
%---------------Previous Works--------------------------------------%
Apart from causing propagation losses and disorder-induced localized
resonances, disorder also induces changes in the eigenfrequencies and
eigenmodes of the underlying PC. Ramunno and Hughes \cite{Ramunno2009}
modelled disorder-induced resonance shifts in PC nanocavities and
predicted a non-trivial disorder-induced mean blueshift in the cavity
resonance. Patterson and Hughes \cite{Patterson2010} extended this
formalism to PCWs, and predicted both a mean blueshift of resonances
and a disorder-induced mode edge broadening. To the best of our
knowledge, this mean blueshift has not been experimentally measured;
this is likely because there was no simple experimental procedure for
proving that a mean blueshift occurs, especially for an intrinsically
disordered PCW.  Both of the theoretical works mentioned above dealt with intrinsic
disorder only, which occurs via rapid fluctuations of the
air-dielectric interface and highlighted the importance of carefully
taking into account local fields at the interface. Recently, Savona
has exploited a guided mode expansion technique to compute
disorder-induced localized modes and the corresponding spectral
density, which, as expected, showed sharp spectral signatures near the
mode edge indicative of spatially localized modes \cite{Savona2011}.
% the ideal bandstructure highlighting the effect of mode mixing
% between the fundamental (even) and higher order (odd) modes.

%------------------Our results------------------------------%
In this paper, we introduce a model to describe disorder-induced
resonance shifts and broadening of the fundamental mode near the
mode-edge which takes into account a systematic increase of the
disorder parameters (i.e., it allows one to model deliberate disorder
which can be controlled and changed in a systematic way).  This leads
to a predictable trend that causes an increasing mean blue shift and a
broadening of the photonic density of states (DOS).  We show how one
can extend the theoretical models introduced in
Refs.~\cite{Ramunno2009,Patterson2010} to account for both intrinsic
and deliberate or extrinsic disorder where extrinsic disorder is
characterized by a deliberate shift of the hole centres. We carefully
include local field effects \cite{Johnson2005} in our model to compute
the first and second-order perturbative changes to the
eigenfrequencies of the fundamental waveguide mode which then allow us
to compute the disordered DOS. For disordered PCWs with varying
extrinsic disorder (see Ref.~\cite{Javadi2014} for details), we
compute the ensemble averaged DOS via a Monte Carlo approach.
Experimentally, measurements of vertically emitted intensity are taken
for GaAs PCW membranes with varying amounts of extrinsic
disorder. Since the intensity measurements are a direct measure of
disordered-induced broadening and frequency shift (blueshift) of the
DOS, we compare our computed DOS with the intensity measurements and
the two are found to be in good qualitative agreement. The comparison
between our theory and experimental data demonstrates the importance
of including local field effects when computing disorder-induced
changes to the eigenfrequencies and eigenmodes of PCWs. While our
theory is perturbative, the semi-analytical approach is
computationally efficient and the results offer useful insights in
designing disordered PCWs for spontaneous emission enhancements of
embedded quantum dots. Finally, we also show an example of the
underlying disorder-induced quasimodes that can be obtained on a
finite-size PC lattice by computing the numerically exact Green
function using a full 3D FDTD approach \cite{Yao2009}.

%who directly measure the disordered-induced
%broadening and frequency shift (blueshift) of the photon density of
%states (DOS) through vertically emitted intensity measurements. 
  
%Indeed many of the current models predict no
%disorder-induced resonance shift at all because of the employed
%polarization model for disorder.
Our paper is organized as follows. In Section~\ref{sec:shifts} we
review our formalism for modelling disorder-induced resonance shifts
and point out the limitations of some of the polarization models
commonly used in literature for modelling disorder in PCWs. We then
introduce our extended polarization model for modelling both intrinsic
and extrinsic disorder and show how it results in a non-vanishing
first-order frequency shift. In Section~\ref{sec:dos}, we highlight
our approach for computing the disordered DOS given disorder-induced
resonance shifts. We also present a mathematical argument based on
photonic Green functions that link the disordered DOS to vertically emitted
intensity measurements. In Section~\ref{sec:results}, we numerically
compute disorder-induced resonance shifts and the disordered DOS which
we compare with experimental measurements performed on GaAs PCW
membranes.  While we find qualitatively good behaviour, in
Section~\ref{sec:ldos} we discuss some limitations of our perturbative
model and we also show some numerically exact simulations of
finite-size PCWs, which are limited in spatial size because of the
numerical complexities. In Section \ref{sec:discussion}, we summarize
the strengths and weaknesses of our perturbative semi-analytic
approach and discuss our results in the context of previous reports in
the literature. We conclude in Section \ref{sec:conclusions}.

%Good qualitative agreement between
%experiment and theory is found.
\section{\label{sec:shifts}Disorder-induced resonance shifts 
and disorder Polarization Models}
For modelling the effects of disorder on light scattering in PCWs, we
focus our attention on deriving the first and second order
perturbative changes to the eigenfrequencies of a dielectric
structure. We treat disorder as a perturbation and employ perturbation
theory techniques adapted to dielectric structures with high index
contrasts \cite{Johnson2002a, Johnson2005}. Denoting the perturbed
eigenfrequencies as
$\omega = \omega_0 + \sum_{i} \Delta \omega^{(i)}$, where
$\omega_0$ is the unperturbed eigenfrequency and $\Delta \omega^{(i)}$
represents the $i^{th}$ order perturbation, we restrict ourselves to
the first $\Delta \omega^{(1)}$ and second order $\Delta \omega^{(2)}$
corrections. Since disorder in PCWs is statistical in nature, we
compute the ensemble average over nominally identical disordered PCWs
denoted by $\EXP{}$. Thus the first-order ensemble averaged correction
is given by \cite{Ramunno2009} ($\omega$ dependence is implicit)
\begin{equation}
  \label{eq:1st_order_shift}
  \EXP{ \Delta\omega^{(1)} } = -\frac{\omega_0}{2}\int_{\mathrm{cell}}
  \EXP{ \mathbf{E}^{*} \left(\mathbf{r}\right) 
    \cdot \mathbf{P}(\mathbf{r}) } d\mathbf{r},
\end{equation}
where $\mathbf{E}(\mathbf r)$ is the unperturbed eigenmode,
$\mathbf{P} (\mathbf r)$ is the polarization function to characterize
the dielectric disorder, and the integration is carried out over the
primitive unit cell of the PC lattice. The fields are normalized
according to $\int_{\mathrm{cell}} \varepsilon(\mathbf r)
\mathbf{E}^{*}\left(\mathbf{r}\right) \cdot
\mathbf{E}\left(\mathbf{r}\right)d\mathbf{r} = 1$, where
$\varepsilon(\mathbf r)$ is the unperturbed dielectric constant. The
ensemble average of the second-order correction $\mathbb{E}[\Delta
\omega^{(2)}]$, is computed similarly \cite{Patterson2010}
\begin{align}
  \label{eq:2nd_order_shift}
   \nonumber & \EXP{ \Delta\omega^{(2)} } = \\
  & \frac{\omega_0^{2}}{4}\iint \EXP{ \mathbf{E}^{*}
  \left(\mathbf{r}\right)\cdot\mathbf{P}\left(\mathbf{r}\right)\mathbf{E}^{*}\left(\mathbf{r}'\right)
  \cdot\mathbf{P} \left(\mathbf{r'}\right) }d\mathbf{r} d\mathbf{r'}.
\end{align}

To interpret the ensemble averages of the first and second order
corrections in a statistical sense, we denote the \emph{net frequency shift}
as $\Delta \omega \left( \sum_{i} \Delta \omega^{(i)} \right),$ with
$\omega = \omega_0 + \Delta \omega.$ The net frequency shift
$\Delta \omega$ is interpreted as a random variable whose probability
distribution has
$\EXP{\Delta \omega^{(1)}},\, \EXP{\Delta \omega^{(2)}}$ as its first
and second-order moments, respectively.
Hence, given the unperturbed eigenmodes, one is left with choosing a
suitable polarization model to describe the perturbation of the PC
lattice. Structural disorder in PCWs can be viewed as introducing
additional scattering sites in an otherwise perfect PCW lattice. The
scattering sites induce dipole moments resulting in a disorder-induced
polarization which acts as a source term in the homogeneous Maxwell
equations, thus contributing to scattering of a propagating Bloch
mode. An alternative picture is that perturbations will disorder the
PC band structure and thus the DOS, which will result in
disorder-induced localization modes and scattering in directions that
might otherwise be forbidden (e.g., a lossless propagating mode will
couple to radiation modes above the light line in the case of a PCW
slab).

There are typically two models that have been widely used in the
photonics community for modelling dielectric perturbations, which
denote as \textit{weak-index contrast}
$\mathbf{P}_{\rm w} (\mathbf r)$ and \textit{smooth-perturbation}
$\mathbf{P}_{\rm s} (\mathbf r)$. The former model neglects the
problem of field-discontinuities at high-index-contrast surfaces, and
is well defined for field components that are parallel to the
interface; in contrast, the latter model addresses this
field-discontinuity problem, though it is appropriate for perturbing a
surface uniformly in a perpendicular direction, e.g., displacing a
long sidewall in a direction that is perpendicular to the wall
interface. By way of a simple example, consider a simple planar
interface between two dielectrics $\varepsilon_a,\, \varepsilon_s$,
located at $\mathbf{r'}$. When perturbed by a small amplitude
$\Delta h$, the two polarization models are given as:
\begin{align}
 & \label{eq:weak_index} \mathbf{P}_{\rm w}(\mathbf r) =   \Delta \varepsilon \Delta h (\mathbf{r})
  \mathbf{E} (\mathbf r) \delta(\mathbf{r}-\mathbf{r'}) , \\
  & \label{eq:smooth} \mathbf{P}_{\rm s}(\mathbf r) =   \Delta \varepsilon \Delta h(\mathbf{r})
  \left( \mathbf{E}_{\parallel}(\mathbf r) +
    \frac{\varepsilon(\mathbf{r})}{\varepsilon_a \epsilon_s}
    \mathbf{D}_{\perp}(\mathbf r) \right) \delta(\mathbf{r}-\mathbf{r'}),
\end{align}
where $\Delta \varepsilon = \varepsilon_a - \varepsilon_s$ or
$\Delta \varepsilon = \varepsilon_s - \varepsilon_a$ depending on the
direction of the perturbation, i.e. from $\varepsilon_a$ to
$\varepsilon_s$ or vice versa, and
$\mathbf{E}_{\parallel}(\mathbf r),\ \mathbf{D}_{\perp}(\mathbf r)$
denote the parallel and perpendicular components of the
electromagnetic fields relative to the boundary interface. The
weak-index contrast model is accurate in systems exhibiting weak-index
contrast (i.e., $|\Delta \varepsilon| \ll 1$) and is the most popular
choice for modelling imperfections in dielectric structures such as
optical waveguides \cite{Marcuse1974}. In high-index contrast systems
such as PCWs, the quantity $\Delta \varepsilon |\mathbf{E}|^2$ is,
however, generally ill-defined at the interface due to a large step
discontinuity in $\mathbf{E}_{\perp}$ \cite{Johnson2002a}, hence the
smooth-perturbation model is likely more appropriate at the interface
due to the use of continuous field components. The smooth-perturbation
model is expected to be valid as long as the perturbation is
smooth. Both models have been used to compute disorder-induced losses
in PCWs \cite{Andreani2007, Wang2008, Hughes2005, Patterson2010} and
have yielded a good qualitative understanding of the observed
disorder-induced loss phenomena.
%Yet it cannot produce a first order shift as we show below.
%Both these models are independent of the
%scatterer's dimensions.

%-------------------Bump Model---------------------------------%
If one views the perturbation (smooth or piecewise smooth) as
introducing scatterers into the system, one must take into account
their respective polarizabilities which in general depend on the
direction of the perturbation. The weak-index contrast and smooth-perturbation models
assign polarizabilities that differ only in sign when the direction of
perturbation is reversed but remain unchanged in magnitude. Moreover,
the magnitude of the polarizability of a scatterer can be drastically
different in the weak-index approximation as demonstrated by the
example of a small dielectric sphere in a homogeneous background (see
Ref.~\cite{Jackson1999}). Therefore, in general for piecewise smooth
perturbations such as bumps on an interface, it is important to
compute polarizabilities that correctly take into account the
direction of perturbation. To address this concern, 
Johnson \emph{et al.}~\cite{Johnson2005} introduced the \textit{bump-perturbation}
polarization model,
 denoted by $\mathbf{P}_{\rm b} (\mathbf r)$ to model
surface roughness in PCWs as piecewise smooth bumps on the interface, where
\begin{align}
  \label{eq:bump}
  \mathbf P_{\rm b}(\mathbf r) =  
   \left[
    \varepsilon_{\mathrm{avg}} \alpha_{\parallel}\mathbf{E}_{\parallel}(\mathbf r) +
    \varepsilon(\mathbf r) \gamma_{\perp}\mathbf{D}_{\perp}(\mathbf
    r) \right] \Delta V \delta(\mathbf{r} - \mathbf{r}'),
\end{align}
where
$\varepsilon_{\mathrm{avg}} = \left(
  \frac{\varepsilon_a+\varepsilon_s}{2} \right)$,
and $\alpha_\parallel,\ \gamma_\perp$ denote the polarizabilities
(polarizability tensors per unit volume) of the bump perturbation and
$\Delta V$ is the volume of the disorder bump element. This model is
valid for arbitrary dielectric contrasts and bump shapes, and useful
formulas have been obtained for rectangular and cylindrical shaped
bumps \cite{Johnson2005}.  Using the polarizabilities for a
cylindrical bump shape, this model has been used to model resonance
shifts caused by intrinsic disorder in PCWs \cite{Patterson2010} where
a mean blueshift and broadening of the ideal bandstructure was
found. As was noted in Ref.~\cite{Patterson2010}, resonance shifts in
the band structure are not predicted by either the weak-index contrast
or the smooth-perturbation models.

\begin{figure}  

  % W1 Schematic
  \begin{tikzpicture}[scale=0.42]    
    % The two for loops create a trianugular PhC (both ideal and disordered). 
    % \y loop controls number of rows. \x loop controls number of holes per row.

    % assigning variables. sets macro/variable. Can't use _ or numbers in name. Parsed by \pgfmathparse
    \pgfmathsetmacro{\sqthree}{sqrt(3)}
    %-----------------Edit to change numrows or holes--------------------------%
    \pgfmathsetmacro{\rows}{2} % acutal num of rows is 2*\rows.
    \pgfmathsetmacro{\numholes}{10} % actual numholes is value-1
    %--------------------------------------------------------------------------%
    
    % Draw rectangular boundary
    \pgfmathparse{\rows+1}
    \fill[white!60!black] (-0.5,-\pgfmathresult*\sqthree) rectangle (\numholes,\rows*\sqthree);
     \node [anchor=south] at (\numholes/2,\rows*\sqthree) {(a)}; %label (a)
   
    % Draw crystal
    \pgfmathparse{\rows-1}
    \foreach \y in {0,...,\pgfmathresult}
    {
      \pgfmathparse{\numholes-1}
      \foreach \x in {0,...,\pgfmathresult}
      {     
        % a = 1
        
        \pgfmathsetmacro{\rad}{0.295}
        % -------------------Edit to change disorder-----------------------------%
        \pgfmathsetmacro{\sigma}{0.1}
        % -----------------------------------------------------------------------%
        % generate disorderd displacement (dx,dy)
        \pgfmathsetmacro{\dx}{(\sigma*rnd)*cos(2*pi*rnd r)} % r is for radians
        \pgfmathsetmacro{\dy}{(\sigma*rnd)*sin(2*pi*rnd r)}       
        
        % Ideal PCW
        % \draw[densely dashed] (\x, \y*\sqthree) circle (\rad); 
        \fill[black!80!white] (\x, \y*\sqthree) circle (\rad); 
        % \draw[densely dashed] (\x + 0.5, \sqthree/2 + \y*\sqthree) circle (\rad);
        \fill[black!80!white] (\x + 0.5, \sqthree/2 + \y*\sqthree) circle (\rad);             
        % \draw[densely dashed] (\x, -\sqthree - \y*\sqthree) circle (\rad); 
        \fill[black!80!white] (\x, -\sqthree - \y*\sqthree) circle (\rad); 
        % \draw[densely dashed] (\x + 0.5, -\sqthree - \sqthree/2 - \y*\sqthree) circle (\rad);
        \fill[black!80!white] (\x + 0.5, -\sqthree - \sqthree/2 - \y*\sqthree) circle (\rad);

        % Disordered PCW
        \draw[solid, color=red, thick] (\x + \dx, \y*\sqthree + \dy) circle (\rad); 
        \draw[solid, color=red, thick] (\x + 0.5 + \dx, \sqthree/2 + \y*\sqthree + \dy) circle (\rad);
        \draw[solid, color=red, thick] (\x + \dx, -\sqthree - \y*\sqthree + \dy) circle (\rad); 
        \draw[solid, color=red, thick] (\x + 0.5 + \dx, -\sqthree - \sqthree/2 - \y*\sqthree + \dy) circle (\rad);
      }     
    }
  \end{tikzpicture}  
  % Circle schematic
  \begin{tikzpicture}[scale=1.3]     
  % Unperturbed circle         
  \draw[dashed] (0,0) circle (1);   
  % Perturbed circle
  \draw[] (45:0.6) circle (1); 
  % Delta r arrow
  \draw[->, thick] (0,0) -- node[anchor=south, xshift=-1.5, yshift=2] {$\Delta r$} (45:0.6) ;
  % Draw the origin
  \draw (-1.5,0)--(1.5,0) (0,-1.5)--(0,1.5);
  \node at (0,1.7) {(b)};
  % The angle arc
  \draw[thick] (0.35,0) arc (0:45:0.35) node[anchor=north, xshift=-2, yshift=1]  {$\phi$};
  % draw unperturbed center
  \draw[fill,orange] (0,0) circle (0.02);
  % draw perturbed center
  \draw[fill,orange] (45:0.6) circle (0.02);

  % arrows showing various points on the circle shifting
  \draw[->, thick, dashed, red] (45:1) -- +(45:0.6);
  \draw[->, thick, dashed, red] (90:1) -- +(45:0.6);
  \draw[->, thick, dashed, red] (135:1) -- +(45:0.6);
  \draw[->, thick, dashed, black] (180:1) -- +(45:0.6);
  \draw[->, thick, dashed, black] (225:1) -- +(45:0.6);
  \draw[->, thick, dashed, black] (270:1) -- +(45:0.6);
  \draw[->, thick, dashed, red] (315:1) -- +(45:0.6);
  \draw[->, thick, dashed, red] (360:1) -- +(45:0.6);

  % draw lines and labels         
  %\draw (0,0)-- node[below] {$r_0$} (30:2.5cm) node[right] {$(x,y)$};      
  %\draw[->,thick] (0:0) -- node[left] {$|\Delta r|$} (140:1.5cm);         
  %\draw[->] node[above] {$\Delta \tilde{\theta}$} (0.6cm,0) arc (0:140:0.6cm);      
  %\draw (140:1.5cm) -- node[above] {$\Delta d$}  (30:2.5cm) ; 
  % draw centre points    
  %\node[inner sep=0.3mm,draw,circle,fill] at (0,0) {};    
  %\node[inner sep=0.3mm,draw,circle,] at (140:1.5cm) {}; 
  %\node[left] at (140:2.5) {$\varepsilon_s$};     
  %\node[right] at (140:2.5) {$\varepsilon_a$}; 
  %\draw[very thick] (130:2.5) arc (130:140:2.5); 
\end{tikzpicture}
%\end{tikzpicture}

%\begin{tikzpicture}     
%  \draw (0,0) -- node[above left] {$\varepsilon_a$} (0,2);         
%  \draw[->] (0,1)-- node[above] {$\varepsilon_s$}(0.5,1);         
%  \draw[dashed] (0.5,0)--(0.5,2); 
%\end{tikzpicture} 
\caption{(Color Online)\label{fig:dis_schematic} (a) Schematic of a
  disordered W1 with ideal air holes (dark-filled/black circles),
  disordered holes (light/red circles) and the background slab
  (grey). (b) Schematic of a hole centre shift with the centres marked
  by light-filled/orange circles and the direction of perturbation
  given by the solid-black arrow. The magnitude of the perturbation is
  denoted by $\Delta r$ and the direction is given by $\phi$. Dashed
  arrows indicate the shifts of the air-slab interface with
  dashed-red/light and dashed-black/dark arrows representing positive
  and negative shifts respectively.}
\end{figure}
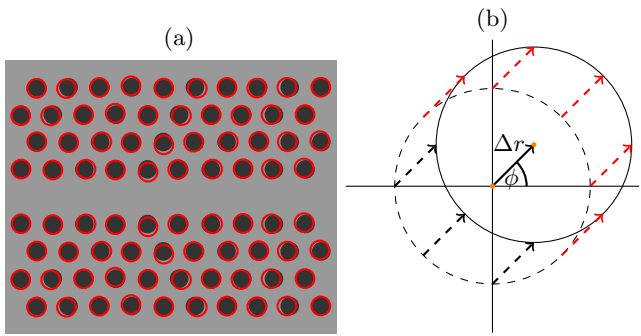

In this work, we apply the \emph{bump-perturbation} model with
cylindrical bump shape polarizabilities, to compute disorder-induced
resonance shifts in PCWs, and systematically investigate what happens
with an increase in the disorder parameters for shifted holes. We use
this model to connect to related experiments on deliberately
disordered GaAs membranes where embedded quantum dots couple to
disorder-induced localized modes resulting in enhanced spontaneous
emission \cite{Sapienza2010, Javadi2014}. The PCW we consider is a
standard W1 formed by introducing a line defect in a triangular
lattice of air holes etched in a semiconductor slab
Fig.~\ref{fig:dis_schematic}(a). The extrinsic disorder perturbation
is characterized by a hole centre shift as shown in
Fig~\ref{fig:dis_schematic}(b). The air holes are cylinders so we
employ cylindrical coordinates $(r,\theta,z)$ henceforth. Furthermore,
the disordered air hole is assumed to have a constant cross section
throughout the slab thickness. This allows us to replace the disorder
volume element in Eq.~(\ref{eq:bump}) by its cross-sectional area
$\Delta A$ and the polarizabilities are now $\mathrm{2\,x\,2}$ tensors
representing polarizability per unit area \cite{Johnson2005}. To first
order, the perturbed area $\Delta A$ of the disorder element is
proportional to $|\Delta h|$ which quantifies the amplitude of the
hole centre shift.
%\com{highlight with ref to fig 1(b) why
%  the smooth polarization model can fail}

%------------Disorder model for Δh-------------------------------%
%The interplay between local field effects and intrinsic disorder was
%previously studied in Ref.~\onlinecite{Patterson2010}, where a mean
%blueshift and broadening of the ideal bandstructure was found by using
%the bump-perturbation model. 
In light of current experiments studying localization modes and
resonance shifts as a function of deliberate disorder, we extend the
disorder model of Ref.~\onlinecite{Patterson2010} to deal with both
intrinsic and varying extrinsic disorder.
%and show how one can compute the disordered DOS.  
While the previous model considered rapid radial fluctuations of the
air-slab interface as the source of intrinsic disorder, here we model
both intrinsic and extrinsic disorder as a net centre shift of the air
hole as shown schematically in Fig.~\ref{fig:dis_schematic}(b).
Although intrinsic disorder is likely best described by rapid radial
fluctuations, the choice to model intrinsic disorder as a hole centre
shift is driven by simplicity as one can map rapid radial fluctuations
to an effective hole centre shift by comparing experimental loss data
with numerical simulations as demonstrated by Garcia \emph{et
  al.}~\cite{Garcia2013}; also, the main effect of the disorder below
is through deliberate disorder.

%----------------------FIGURE-----------------------------------%
%\input{diswguide_schematic}
%---------------------------------------------------------------%
Since disorder is stochastic in nature, we denote $(\Delta h,\, \phi)$
as the random variables quantifying the total disorder (extrinsic and
intrinsic) in PCWs. The net amplitude of the shift $|\Delta h|$ is
constant around the circumference while the sign is determined by the
net azimuthal direction of the shift, denoted by $\phi$. The shift of
an infinitesimally small arc lying on the circular air-slab interface
is then given by
\begin{equation}
  \label{eq:total_disorder}
  \Delta h (\Delta r, \phi; \theta) = 
  \begin{cases}
    + \Delta r, \: \phi \in \Omega , \\
    - \Delta r, \: \phi \notin \Omega,
    %[\theta - \frac{\pi}{2}, \theta + \frac{\pi}{2}]
  \end{cases}
\end{equation}
where $\Omega = [\theta - \frac{\pi}{2}, \theta + \frac{\pi}{2}]$,
$\theta$ denotes the polar coordinate and $\Delta r$ quantifies the
net radial perturbation. A positive bump/shift ($+\Delta r$) is defined as
the air-slab boundary shifting into the slab and vice versa for
negative bumps/shifts as illustrated in
Fig.~\ref{fig:dis_schematic}(b). We denote $\Delta r_{i/e} \sim
|\mathcal{N}(0,\sigma_{i/e})|,\ \phi_{i/e} \sim \mathcal{U}[-\pi,\pi]$
as the random variables for the radial magnitudes and azimuthal
directions of the intrinsic/extrinsic disorder perturbations,
respectively;  $\mathcal{N}(\mu,\sigma)$ denotes a normal
distribution with mean $\mu$ and standard deviation $\sigma$ while
$\mathcal{U}[a,b]$ denotes a uniform distribution on the interval
$[a,b]$. The net radial fluctuation can be broken down into its
Cartesian components $\Delta x,\, \Delta y$,  which are given below
\begin{align}
  \label{eq:total_disorder_cartesian_components}
  \Delta x &= \Delta r_i \cos(\phi_i) + \Delta r_e \cos(\phi_e), \\
  \Delta y &= \Delta r_i \sin(\phi_i) + \Delta r_e \sin(\phi_e).
\end{align}
The net radial fluctuation is then given as
$\Delta r = \sqrt{\Delta x^2 + \Delta y^2}$ while the net azimuthal
direction is simply
$\phi = \tan^{-1}\left(\frac{\Delta y}{\Delta x} \right)$.  Comparing
to our previous model of rapid radial
fluctuations~\cite{Patterson2010}, this model lacks the concept of a
intra-hole correlation length as all points on the hole shift by the
same magnitude but in different directions depending on the angular
hole coordinate $\theta$. However this model is more appropriate for
modelling the deliberate displacement of the disordered holes
performed in the experiment.

%-------------First order expectation for diff models------------------------% 
%-------------Weak and Slowly Varying Models------------------------%
To highlight the main difference between the three polarization models
discussed earlier, let us look at computing the ensemble averaged
first-order frequency shift $\mathbb{E}[\Delta \omega^{(1)}]$ by using
Eqs.~(\ref{eq:weak_index}), (\ref{eq:smooth}) or (\ref{eq:bump}) in
Eq.~(\ref{eq:1st_order_shift}). For weak-index contrast and
smooth-perturbation models, one must compute the expectation of the
total disorder $\EXP{\Delta h}$. If the extrinsic disorder is zero
($\sigma_e=0$), it is trivial to show that $\EXP{\Delta h} = 0 \to
\mathbb{E}[\Delta \omega^{(1)}] = 0$. In the case where both intrinsic
and extrinsic disorder are present, one can still show
$\mathbb{E}[\Delta h] = 0$ as can be verified via a Monte Carlo
simulation. This result is expected because given any random value for
the net radial displacement $\Delta r$, all possible azimuthal
directions are equally likely; and since we are assigning symmetric
weights (differing only is sign) to positive and negative shifts in
these two models, the first order correction vanishes.
%Therefore for weak-index contrast and
%smooth-perturbation polarization models, the net mean frequency shift
%is zero. 
This is in line with  previous findings where intrinsic disorder was modelled as rapid radial fluctuations
\cite{Patterson2010}.
%--------------Cylindrical bump model----------------------------%
However, a non-zero first-order mean frequency shift is expected to
occur for the bump-perturbation model since
$\mathbb{E}[\alpha_\parallel\Delta h] \neq 0,\,
\mathbb{E}[\gamma_\perp\Delta h] \neq 0$.
This is because the polarizabilities for the shifts that we use in
Eq.~(\ref{eq:bump}) are {\em asymmetric}, i.e.
$\alpha_{\parallel}^+ \neq \alpha_{\parallel}^-,\, \gamma_{\perp}^+
\neq \gamma_{\perp}^-$
where $+/-$ denote positive and negative shifts respectively.
%and this is responsible for a
%non vanishing first-order frequency shift ($\mathbb{E} \left[
 % \Delta\omega \right] \neq 0$). 
For the second order correction $\EXP{\Delta \omega^{(2)}}$, none of
the expectation terms vanish and therefore the variance of the net
mean frequency shift is non zero for all three polarizability models.
One way to test which model is more appropriate is to compare with
experiments where the amount of disorder can be controlled, and that
is precisely what we do below.

\section{\label{sec:dos}Disordered density of states and 
  connection to experiments of vertical light emission}
Since disorder acts to shift and broaden the mode edge, a useful
quantity for experimental comparison is the DOS
$\rho (\omega(\mathbf k))$, defined as the number of frequency levels
per unit volume of ${\bf k}$-space. Unlike the concept of
bandstructure, which is only well-defined in perfectly periodic
systems, the DOS is valid for all structures. It is well known that
the DOS of an ideal PCW diverges at the mode edge since the group
velocity vanishes, while for a disorder PC structure, the ensemble
averaged DOS exhibits a broadened peak around the ideal mode edge
where the width of the peak is proportional to the amount of disorder
present in the PC structure \cite{Fussell2008}.
%Hence the signature of a mode edge
%is a brodened peak in the DOS/LDOS. 

% -----------Numerics of DOS------------------------------------------%
To compute the DOS, we first remark that the definition of DOS bears
close resemblance to the mathematical definition of a probability
density function (PDF). Hence, just like a histogram generated from a
large sample dataset represents the underlying PDF, the histogram
generated from a bandstructure represents the DOS. To generate a
disordered DOS instance, we generate a disordered bandstructure given
by $\omega(k) = \omega_0(k) + \Delta \omega(k)$ where the net
frequency shift $\Delta \omega(k)$ is a random variable \emph{assumed}
to have a normal distribution with mean
$\EXP{\Delta \omega^{(1)} (k)}$ and variance
$\EXP{\Delta \omega^{(2)}(k)}$. The disordered bandstructure allows us
to calculate an instance of the disordered DOS $\rho(\omega)$. One
then computes the ensemble averaged disordered DOS
$\bar{\rho}(\omega)$ by averaging over many such disordered DOS
instances.

%\com{this is a bit rough for the moment - just for discussion:}
Experimentally, the DOS can be obtained by spatially averaging the
vertically emitted light intensity measurements in PCWs. To appreciate
how the waveguide DOS can be measured through vertical emission,
consider the waveguide mode Green function without any
disorder~\cite{RaoPRB2007}:
\begin{align}
{\bf G}_{\rm wg}({\bf r},{\bf r}';\omega)
&= \frac{ia\omega}{2v_g}
[\Theta(x-x') {\bf f}_k({\bf r})  {\bf f}_k^*({\bf r}') e^{ik(x-x')}
\nonumber \\
&+ \Theta(x-x') {\bf f}_k^*({\bf r}')  {\bf f}_k({\bf r}) e^{-ik(x-x')}],
\end{align}
where $\Theta(x-x')$ is the Heaviside function, $k=k(\omega)$ and
$\mathbf{f}_k$ is the ideal Bloch mode.  Now consider adding a point
disorder model, where the disorder causes a polarizability with a
Lorentzian lineshape (e.g., typical of a disorder-induced resonance or
an embedded light source such as a quantum dot),
$\alpha^{\rm d} = A/(\omega_0-\omega-i\gamma)$, where $\omega_0$ is
the disorder induced resonance frequency, $\gamma$ is the broadening
of the resonance, and $A$ is the coupling strength. Using a Dyson
equation,
$\tilde {\bf G} = {\bf G} + {\bf G} \alpha^{\rm d} \tilde{\bf G}$, the
Green function in the presence of the perturbation can be exactly
obtained through $\tilde {\bf G}$, where $\alpha^{\rm d}$ has units of
volume (polarizability volume) and the Green function has units of
inverse volume.  Defining
$\rho_i({\bf r}_{\rm d}) \equiv {\rm Im}[{\bf G}_{ii}({\bf r}_{\rm
  d},{\bf r}_{\rm d})]$
where $\mathrm{Im}[]$ denotes the imaginary component as a measure of
the (projected) LDOS, we obtain
%(\com{check my calculation)}
\begin{align}
\tilde \rho_i({\bf r}_{\rm d},\omega)=
\rho_i({\bf r}_{\rm d},\omega) \frac{(\omega-\omega_0)^2+\gamma[\gamma+
A\rho_i({\bf r}_{\rm d},\omega)]}{(\omega-\omega_0)^2 +[\gamma+A\rho_i({\bf r}_{\rm d},\omega)]^2},
\end{align}
and thus the disordered LDOS now contains signatures of the original
waveguide LDOS and the underlying resonance of the disorder site.
Looking at the limit $\omega\rightarrow\omega_0$, 
\begin{align}
\tilde \rho_i({\bf r}_{\rm d},\omega_0)=
\rho_i({\bf r}_{\rm d},\omega_0) \frac{\gamma}{\gamma+A\rho_i({\bf r}_{\rm d},\omega_0)},
\end{align}
where we note that the LDOS at the mode edge is no longer divergent,
and instead $\tilde{\rho}_i({\bf r}_{\rm d},\omega_{\rm e})=\gamma/A$
(assuming $\omega_0=\omega_{\rm e}$), which is simply the LDOS from
the disordered polarizability model. Since this disordered LDOS is now
connected to light propagation away from the waveguide, vertically
emitted light will clearly contain signatures of the disordered LDOS
for the waveguide modes, and thus the disordered DOS when spatially
integrated.

%When cavity resonances are coupled to PWC modes
%Note models for cavities coupled to PCW modes also show that the
%cavity modes and waveguides modes become intertwined where this
%coupling is solved analytically \cite{YaoPRB2009}.  
An alternative picture of the disordered DOS can be obtained by
connecting directly to a sum over the disordered-induced modes.  In
PCWs, propagating and localized modes couple with radiation modes
above the light line resulting in vertically emitted intensity. Near
the mode edge, the DOS increases due to vanishing group velocity
leading to an increase in the radiation loss rate and a broadened peak
in the vertically emitted intensity spectrum. Other peaks in the
spectrum near the mode edge indicate the presence of disorder-induced
localized modes.  Given that the vertically emitted intensity is
proportional to the radiation loss rate, denoted by $\gamma$, one can
show that the radiation loss rate is proportional to the DOS. Let us
assume the disordered quasimodes (or quasinormal modes)
\cite{Kristensen2014} are known or can be computed, denoted by
$\mathbf{\tilde{f}}_{j}(\mathbf{r})$ where $j$ indexes the quasimodes
which have complex eigenfrequencies
$\tilde\omega_j =\omega_j+i\gamma_{j}$, where the quality factor of
each resonance is $Q_j=\omega_j/2\gamma_j$. Then using mode expansion,
one obtains the Green function of the disordered PCW by
\cite{Lee1999,Ge2014}
\begin{align}
  \label{eq:gf_dis}
  {\mathbf{G}}_{\mathrm{dis}}(\mathbf r, \mathbf r'; \omega) = 
  \sum_j \frac{\omega^2}{2\tilde \omega_j(\tilde\omega - \omega)} 
  \mathbf{\tilde{\bf f}}_{j}(\mathbf{r})  \mathbf{\tilde{\bf f}}_{j}^*(\mathbf r'),
\end{align}
and the  LDOS of
the disordered PCW is 
\begin{align}
  \label{eq:ldos_total}
  \rho (\mathbf{r}, \omega) = \frac{2}{\pi \omega} \mathrm{Im}\left[ 
    \mathrm{Tr} \lbrace {\mathbf{G}}_{\mathrm{dis}}(\mathbf r, \mathbf r;\omega) \rbrace  \right],
\end{align}
where $\mathrm{Tr}[]$ denotes the trace. From the total LDOS one can compute
the DOS by integrating over all space. Therefore one sees that the
radiation loss rate and the vertically emitted intensity is inherently
linked to the disordered DOS. Indeed, each one of the underlying
quasimodes (and every disordered element) has a vertical decay channel
associated with vertical decay above the light line.

%What are the implications of this? It
%seems, experimentally, your samples must have a disorder-induced
%polarization that breaks symmetry and so models $P_w$ and $P_s$ cannot
%be used (though they are commonly used by the community).

%$\alpha_\parallel = \alpha_\parallel
%( \mathrm{sgn} (\Delta h)$ is dependent on the sign of $\Delta h$ and
%$\alpha_\parallel^+ \neq \alpha_\parallel^-$. The same holds for gamma
%and $\alpha_\parallel \neq \gamma_\perp$

%$r_{i}$ denotes all points on the circumference of the
%$i^{th}$ hole and The sum is over the number of holes in the unit
%cell.  Strictly speaking, $\Delta h_i$ is dependent on the $i^{th}$
%hole.  
%where $\varepsilon_{a,s}$ denote the dielectric constants of air and
%the slab respectively. 
%The sum is over the number of holes in the unit
%cell, $r_{i}$ denotes all points on the circumference of the $i^{th}$
%hole, $H$ is the Heaviside function restricting us to within the PC slab which
%is assumed to be of height $h$, the Dirac delta function restricts us
%to the circumference of the $i^{th}$ hole and $\Delta h_{i}(r,\theta)$
%(assumed to be independent of $z$) is a term quantifying the perturbation at
%the circumference.

\section{\label{sec:results}Calculations and measurements of the 
  disordered-induced resonance shifts and  Density of States}
%-------------------------Experimental Parameters----------------------------%
The experimental samples are W1 GaAs membranes with a pitch of
$a=240\,\mathrm{nm}$ and thickness \SI{150}{\nano\metre} with an
embedded layer of InAs self-assembled quantum dots at the centre of
the membrane having uniform density of
\SI{80}{\micro\metre^{-2}}. Quantum dots present a very similar
refractive index to that of the surrounding membrane material
(GaAs). In addition, the experiments presented in this paper are
carried out under high excitation power
(\SI{57}{\micro\watt\per\meter\squared}) \cite{Javadi2014}, which
drives the quantum dots beyond saturation, and they become
transparent. For these reasons, we consider negligible quantum dot
contribution to both inelastic and elastic scattering and, thus, we
can rule out any quantum dot contribution to our model.
% \com{layer of
  % QDS needs to be discussed and as a discussion about how their effect
  % can approximately be neglected altogether - not clear to me why -
  % surely these are also a source for disorder, but hopefully as
  % frequencies outside the measurements because of???}
Various samples each measuring \SI{100}{\micro\metre} long are
manufactured with varying degrees of extrinsic disorder. Extrinsic
disorder is introduced via an additional hole centre displacement
characterized by $\sigma_{\rm e}$ and is varied from
$0.01a = \SI{2.4}{\nano\metre} $ to $0.05a=\SI{12}{\nano\metre}$ in
$0.01a=\SI{2.4}{\nano\metre}$ steps. The samples are excited and
vertically emitted intensity is collected as function of wavelength
and position along the waveguide direction $I(\lambda,x)$ as shown in
Figs.~\ref{fig:dos}(a,b). The intensity is then spatially integrated
along the waveguide $I(\lambda) = \int I dx$ as shown in
Fig.~\ref{fig:dos}(c).

\begin{figure}[] 
  \begin{tikzpicture}
    \matrix[inner sep=0mm]{
      \node [xshift=-1mm](bs) {\includegraphics[scale=0.28]{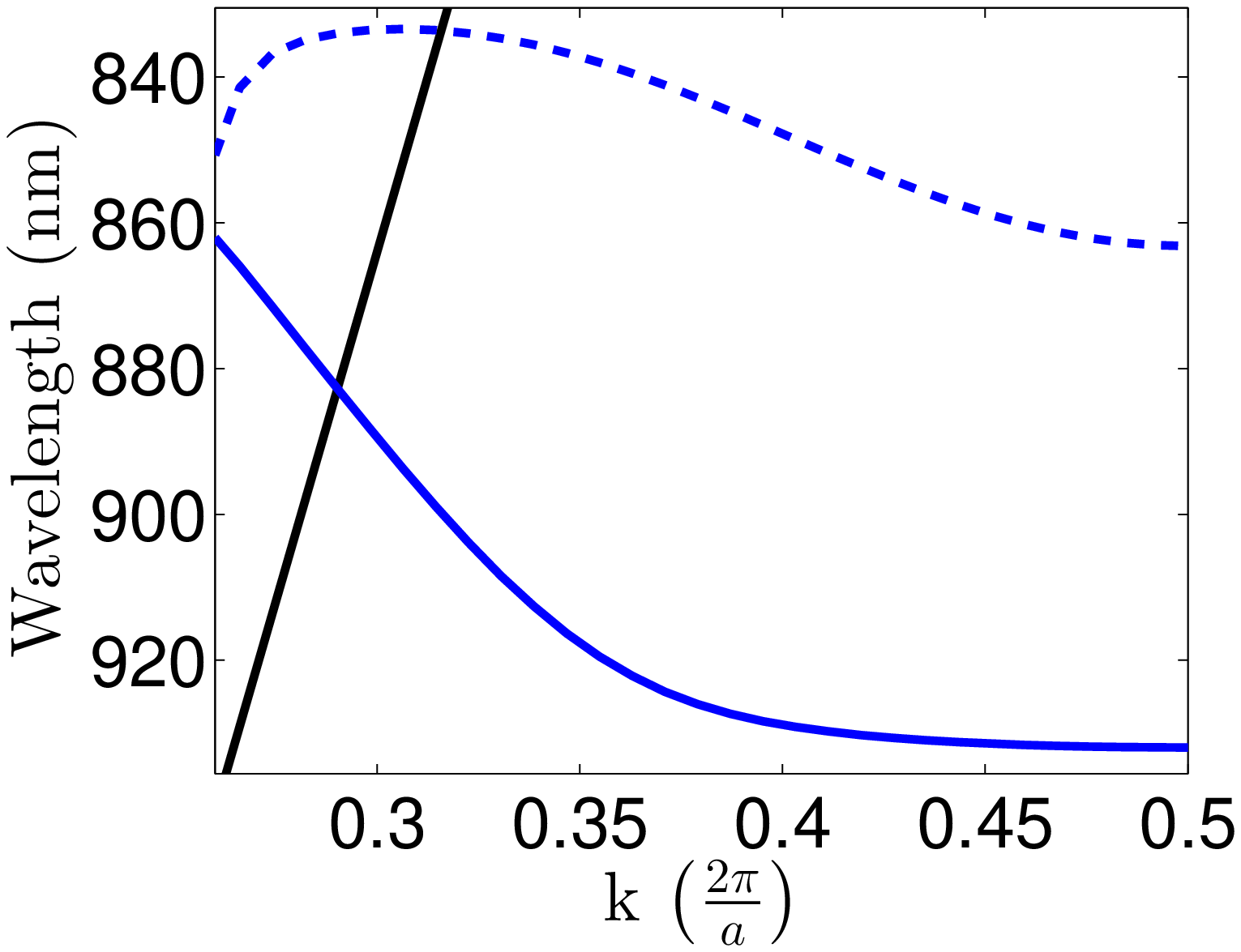}}; &
      \node [xshift=2mm] (dis_bs) {\includegraphics[scale=0.28]{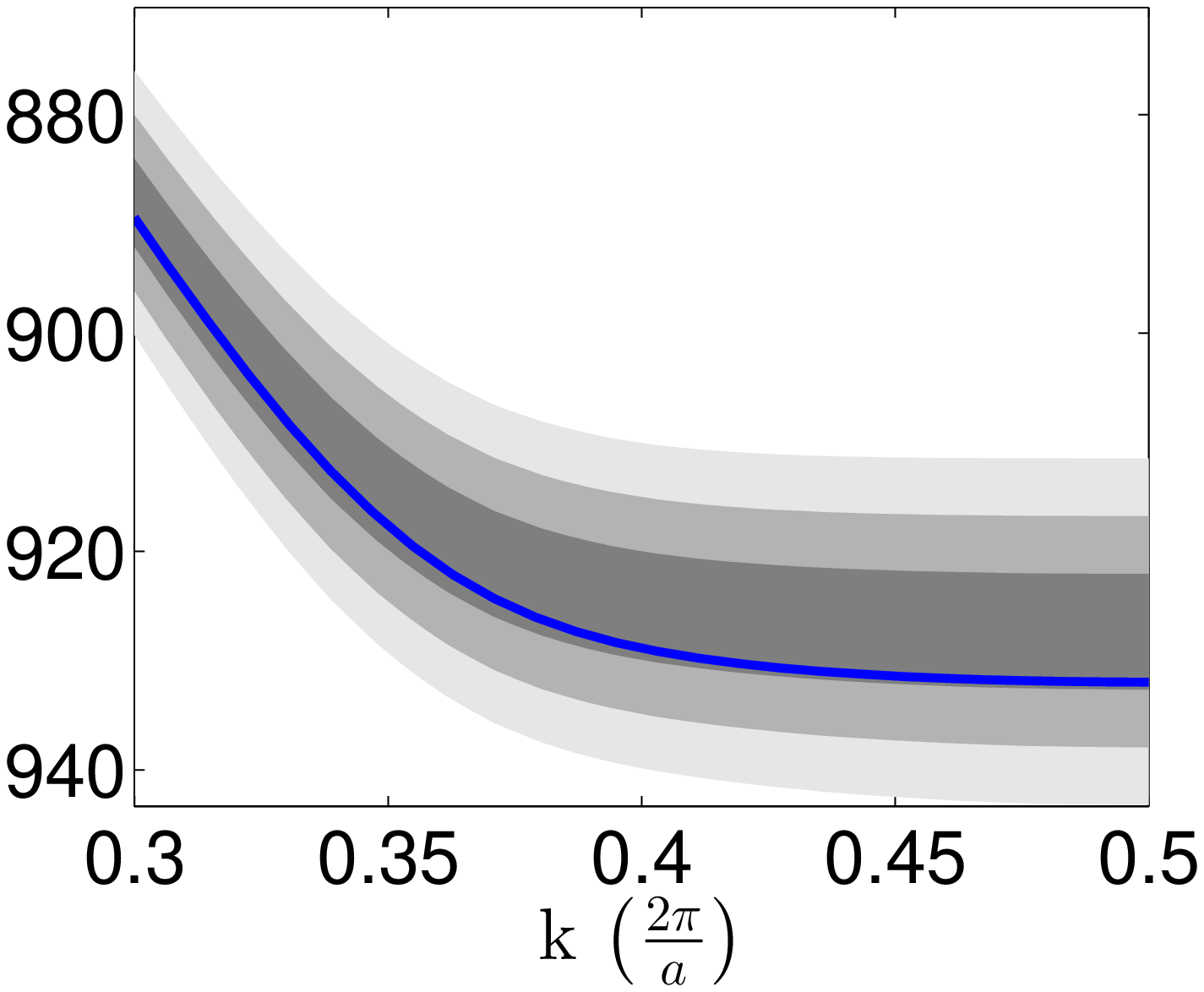}}; \\
      \node[xshift=-2mm] (dlambda) {\includegraphics[scale=0.275]{{{fig_dlambda_1.2_effective}}}}; &
      \node (dlambda_stdev) {\includegraphics[scale=0.275]{{{fig_dlambdastdev_1.2_effective}}}};\\
    };
    \node[anchor=north east, xshift=-5mm] at (bs.north east) {(a)};
    \node[anchor=north east, xshift=-8mm] at (dis_bs.north east) {(b)};
    \node[anchor=south west, xshift=8mm, yshift=5mm] at (dlambda.south west) {(c)};
    \node[anchor=north east, xshift=-8mm] at (dlambda_stdev.north east) {(d)};
  \end{tikzpicture}
  \caption{(Color online) \label{fig:bs_and_shifts}(a) Photonic bandstructure
    (in units of wave vector versus vacuum wavelength) of the ideal W1
    showing the fundamental (solid) and higher order (dashed) guided
    modes. The light line is shown in black.  (b) The broadened
    (blueshifted on average) bandstructure of the unperturbed fundamental mode
    (blue/dark-solid line) for extrinsic disorder of
    $0.02a(\SI{4.8}{\nano\metre})$. The three greyscale shades
      indicate the statistical distribution of the perturbed
      eigenfrequencies that lie within
      $\pm \sigma,\, \pm2\sigma,\, \pm3\sigma$ (dark-grey,
      medium-grey, light-grey) of the unperturbed fundamental mode, where $\sigma$ denotes the standard
      deviation. (c) Ensemble averaged
      first-order eigenvalue corrections for six disordered samples
      representing the mean net frequency shift.
    (d) Ensemble averaged second-order eigenvalue corrections
    representing the standard deviation of the mean net frequency
    shift. In both graphs, the
    \emph{intrinsic} disorder is kept fixed at
    \textcolor{cyan}{$0.005a(\SI{1.2}{\nano\metre})$} while the
    external disorder is varied as follows:
    \textcolor{cyan}{$0a$} (cyan/solid-light-grey),
    \textcolor{green}{$0.01a(\SI{2.4}{\nano\metre})$} (green/dashed-light-grey),
    \textcolor{red}{$0.02a(\SI{4.8}{\nano\metre})$} (red/solid-medium-grey),
    \textcolor{magenta}{$0.03a(\SI{7.2}{\nano\metre})$} (magenta/dashed-medium-grey),
    \textcolor{blue}{$0.04a(\SI{9.6}{\nano\metre})$} (blue/solid-dark-grey),
    \textcolor{black}{$0.05a(\SI{12}{\nano\metre})$} (dashed-black).
    % $0,\, 0.01a=2.4\,\mathrm{nm},\,
    % 0.02a=4.8\,\mathrm{nm},\, 0.03a=7.2\,\mathrm{nm},\,
    % 0.04a=9.6\,\mathrm{nm},\, 0.05a=12\,\mathrm{nm}$. 
  }
\end{figure}

To connect to these experiments, we model a corresponding W1 PCW (see
Fig.~\ref{fig:dis_schematic}(a)) with a slab dielectric constant
suitable for GaAs ($\varepsilon = 12.11$), with the following
parameters: $r=0.295a$ (hole radius), $h=0.625a$ (slab height). The
ideal bandstructure (i.e., with no disorder) is plotted in
Fig.~\ref{fig:bs_and_shifts}(a), depicting the fundamental lossless
guided mode that spans from \SI{876}{\nano\metre} (thereafter going
above the light line) to \SI{930}{\nano\metre} (mode edge). The
intrinsic disorder is approximated as an effective hole centre of
shift (see Ref.~\cite{Garcia2013}), and is kept fixed at
$\sigma_{i}=0.005a=\SI{1.2}{\nano\metre}$.

%------------Computation of shifts--------------------%
The ensemble averaged first and second-order frequency corrections
assuming the bump-perturbation model 
%(the bump is assumed to be of a
%cylindrical shape with the polarizabilities given in
%Ref.~\cite{Johnson2005}) 
for all samples are plotted in Figs.~\ref{fig:bs_and_shifts}(c),
(d). The expectations in
Eqs.~(\ref{eq:1st_order_shift},\ref{eq:2nd_order_shift}) were computed
numerically, that is given the statistical parameters for disorder,
$N$ samples of the set
$(\Delta r_i,\, \Delta r_e,\, \phi_i,\, \phi_e)$ are drawn from the
underlying probability distributions which yields $N=10^4$ samples of
$\Delta h$ . 
%We use $N=10^4$ samples and the numerical mean is then
%taken to approximate the expectations. 
The integration is carried out via Riemann sums where the step size is
chosen to be small enough (\SI{3}{\nano\metre}) to ensure numerical
convergence.
%Given disorder
%parameters, a large number of instances ($\approx 10000$) of the total
%disorder $\Delta h$ are sampled from the underlying probability
%distributions mentioned in the previous section.
%One then takes the mean of these instances to get an
%approximation for the expectation value. 

%-------------------------------FIGURE------------------------------------%
%--------------------------------------------------------------------------%

%--------------------Discuss shifts------------------------------------%
Interpreting the first and second order frequency corrections as the
mean and variance of the net frequency shift (see
Sec.~\ref{sec:shifts}), from Fig.~\ref{fig:bs_and_shifts}(c,d), we see
that for all cases of disorder, the mean frequency shift is a
\emph{blueshift} with a standard deviation that increases as the total
amount of disorder increases. We note that the prediction of a mean
blueshift is non-trivial and is solely due to the asymmetric
polarizabilities present in bump-perturbation polarization
model. Since one is often concerned with the mode-edge or cutoff of
the fundamental guided mode in PCWs, we find that the mode-edge mean
shift and variance is roughly of the same order as the amount of
disorder in the system. For example, for extrinsic disorder of
$0.02a$(\SI{4.8}{\nano\metre}), the mode edge is blueshifted roughly
by \SI{4}{\nano\metre} with a standard deviation of approximately
\SI{4}{\nano\metre}. From a bandstructure point of view, the mean
frequency shift and variance act to shift and broaden the
bandstructure as shown in Fig.~\ref{fig:bs_and_shifts}(b) for the case
with $0.02a$(\SI{4.8}{\nano\metre}) of extrinsic disorder. The three
grayscale shades demonstrate the statistical nature of the frequency
shift by differentiating between frequencies that lie within
$\omega_0 \pm \sigma,\, \omega_0 \pm 2\sigma,\, \omega_0 \pm 3\sigma$
where $\sigma$ denotes the standard deviation and $\omega_0$ is the
unperturbed frequency.
%given by $\sqrt{\EXP{\Delta \lambda ^{(2)}}}$.

%\com{We also need to discuss these trends and convincingly explain them to also reinforce the idea that the numerics are working.}

%-------------DOS Results and Numerics------------------------------------%
The normalized experimental intensity spectra for two different
amounts of extrinsic disorder along the waveguide is shown in
Figs.~\ref{fig:dos}(a,b). Integrating along the waveguide direction,
the corresponding intensity spectra is compared to the ensemble
averaged disordered DOS for the six samples (considered previously in
Fig.~\ref{fig:bs_and_shifts}) in Figs.~\ref{fig:dos}(c,d). For now we
neglect the contribution of radiation modes to the DOS which scales
roughly as $1/\lambda^2$ (Please see Fig.~\ref{fig:dos}(c) for how
this might look like).  Treating the DOS as a probability distribution
as mentioned in Sec.~\ref{sec:dos}, each DOS instance histogram had a
sample size of 1000 (number of k-points) and bin resolution of
\SI{0.27}{\nano\metre} (200 bins). The ensemble-averaged disordered
DOS was calculated from 500 DOS instances. Note, as discussed earlier,
the DOS at the mode edge formally diverges (as the group velocity
approaches zero) in the absence of disorder but our computed
disordered DOS is non divergent and shows a pronounced mean blueshift
as well as broadening caused by the variance of the net frequency
shift. This agrees qualitatively well with the experimental intensity
spectra except for the cases of high extrinsic disorder
$\sigma_{\rm e}=0.05a$(\SI{12}{\nano\metre}) where the theory
overestimates the broadening and for
$\sigma_{\rm e}=0.02a$(\SI{4.8}{\nano\metre}); where the theory
predicts a blueshift, but the experimental intensity spectra is
redshifted. The observed redshift of the mode-edge is within the
computed variance so it is either that this discrepancy arises due to
the experimental sample representing only one disorder instance or the
fabrication method of these particular waveguides where a, e.g.,
proximity effect could introduce an additional unknown degree of
disorder different from the designed one.

We now discuss some limitations to our perturbative approach. For
extrinsic disorder values greater than or equal to
$0.04a$(\SI{9.6}{\nano\metre}), the computed DOS is too broad when
compared to the measured intensity, see
Fig.~\ref{fig:dos}(c),(d). This broadening results from the increase
in standard deviation of mean frequency shift as shown in
Fig.~\ref{fig:bs_and_shifts}(d). Strictly speaking, our perturbation
theory computes mode edge resonance shifts and broadening for
\emph{periodically} disordered PCWs; that is the primitive unit cell
is disordered and then repeated indefinitely. This is an approximation
as in reality the disordered PCW is a concatenation of disordered unit
cell instances sampled from an underlying probability
distribution. Moreover for the extreme extrinsic disorder case of
$0.05a$(\SI{12}{\nano\metre}), we can see the signature of new
localized mode forming below the mode edge around
\SI{945}{\nano\metre} in the intensity spectrum
(Fig.~\ref{fig:dos}(c)) which our computed DOS cannot reproduce since
localized modes that form due to cavity-like defects are naturally not
present in a waveguide exhibiting periodic disorder.
% Moreover, our perturbative approach only computes disorder-induced
% resonance shifts and broadening for frequencies of the fundamental
% mode.

%Multiple scattering becomes important in the slow-light regime
%\cite{Patterson2009}. 
To assess the role of multiple scattering qualitatively, we considered
incoherent disorder-induced losses in our samples, with and without
multiple scattering. With the mode edge roughly corresponding to a
group index of $n_g \approx 50$, our computations indicate that for
$n_g>20$, we are already in the regime of multiple scattering for all
amounts of disorder. Therefore, akin to the overestimation of losses
without multiple scattering \cite{Patterson2009}, the \emph{periodic}
disorder perturbative approach provides an upper bound for mode edge
broadening and for more realistic predictions, a nonperturbative
approach is needed that takes into account multiple scattering
effects. Such an approach is very numerically demanding and is beyond
the scope of this first paper on the topic. However, below we show
some numerically exact solutions of disorder-induced resonances and
LDOS for short length PCWs.
%but a non-perturbative multiple scattering approach is needed to properly
%account for broadening and more importantly localized modes.  

\section{\label{sec:ldos}Numerically Computed Disordered Instances from a finite-size PCW}
% We stress the importance of using the cylindrical bump
% polarization model in computing a mean blueshift; the widely used weak
% contrast or slowly varying surface polarization models employed in the
% literature (see discussions in \cite{Patterson2010}) give a zero mean
% frequency shift and typically use the ideal (non-disordered) DOS.

%Our results prove that disorder not only causes propagation losses, but %it also has a significant effect on the PWC bandstructure
%and DOS.
%---this is because the cylindrical bump model treats positive
%and negative bumps asymmetrically.
% while the other two models treat
%them symmetrically.

%The spectra is obtained by out of plane loss. A probe is scanned along
%waveguide for intensity. Denote $x$ as waveguide direction and
%$\omega$ as frequency. The experimenatlists obtain $I(x,\omega)$ which is then
%integrated along x to give the integrated spectrum. This is somehow close
\begin{figure}
  \begin{tikzpicture}[]
    %[inner sep=0mm]
    % graph for DOS
    % yshift is for shifting the node in y coordinate from the standard position
    %\node [anchor=south west] (fig1a) {\includegraphics[width=0.5\columnwidth]{{{./figures/fig_dosdis_normalized_lambda_1000_200_150_1.2_effective}}}};
    %\node [anchor=south west] (fig1a) {\includegraphics[width=0.5\columnwidth, height=0.53\columnwidth]{{{./figures/fig_dosdis_normalized_gaussian_lambda_1000_200_10000_1.2_effective}}}};
    % labels for the figure
    %\node [yshift=12mm] [below=of fig1a] (labela) {(a)}; % label (a)
    % graph for intensity
    %\node [xshift=-11mm] (fig1b) [right=of fig1a] {\includegraphics[width=0.5\columnwidth, height=0.53\columnwidth]{./figures/fig_exp_data}};
    %\node [yshift=12mm] [below=of fig1b] (labelb) {(b)}; % label (a)

    \matrix[inner sep=0mm] (figs) {
      \node [xshift=0mm](raw_int0) {\includegraphics[scale=0.36]{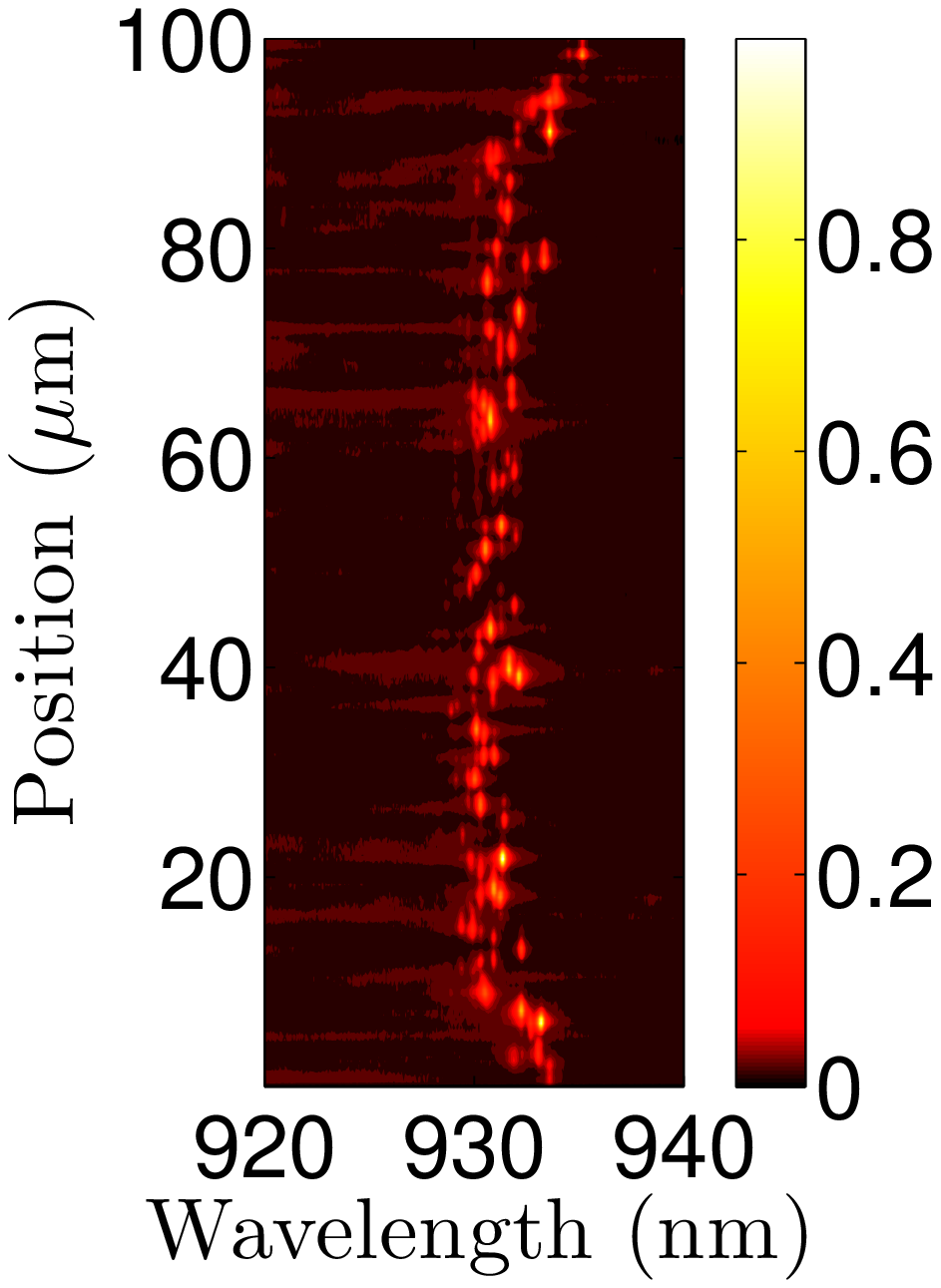}}; &
      \node [xshift=0mm] (raw_int1) { \includegraphics[scale=0.36]{{{fig_exp_data_int_vs_pos_sigma_e_0.03}}} };\\
      \node (exp_integrated_int) {\includegraphics[width=0.51\columnwidth]{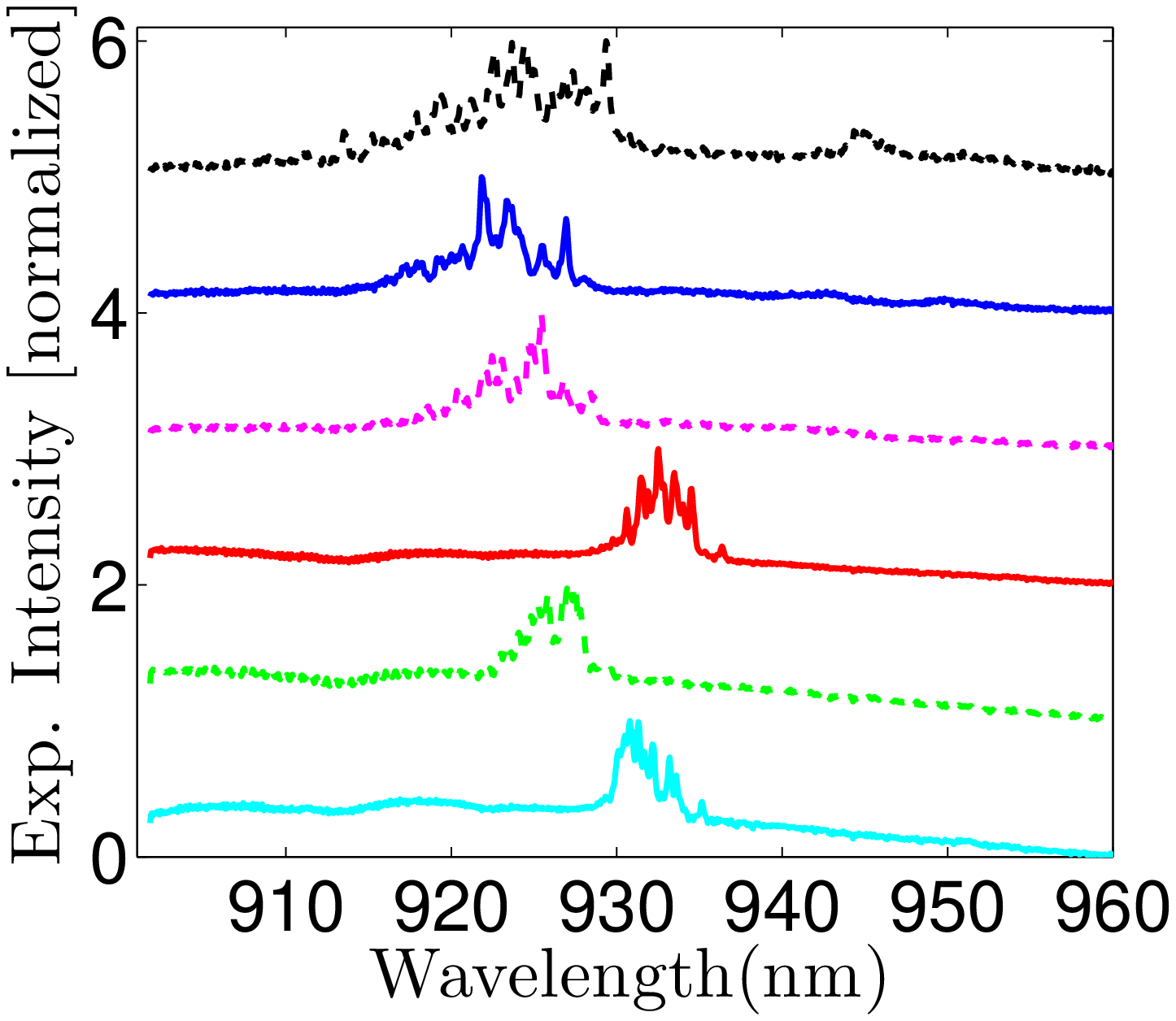}}; &
      \node[xshift=-3mm] (dosdis_normalized) {\includegraphics[width=0.51\columnwidth]{{{fig_dosdis_normalized_gaussian_lambda_1000_200_10000_1.2_effective}}}};\\
    };

    % labels for dos_dis_normalized without matrix env
    %\begin{scope}[x={(dosdis_normalized.south east)}, y={(dosdis_normalized.north west)}]

      % without matrix env
      %\node at (0.5,0.9) {$0a$};
      %\node at (0.83, 0.35) {$0.01a$};
      %\node at (0.83, 0.46) {$0.02a$};
      %\node at (0.83, 0.58) {$0.03a$};
      %\node at (0.83, 0.72) {$0.04a$};
      %\node at (0.83, 0.83) {$0.05a$};
   %end{scope}

   % In matrix env, the node dosdis_normalized has (x,y) flipped.            
    \begin{scope}[y={(exp_integrated_int.south east)}, x={(exp_integrated_int.north west)}]
      \node[cyan] at (0.15, 0.8) {$0a$};       
      \node[green] at (0.15, 0.68) {$0.01a$};       
      \node[red] at (0.15, 0.56) {$0.02a$};       
      \node[magenta] at (0.15, 0.44) {$0.03a$};       
      \node[blue] at (0.15, 0.32) {$0.04a$};       
      \node[black] at (0.15, 0.20) {$0.05a$};       
    \end{scope}
    
    \node[cyan] at (raw_int0.north) {(a) $0a$};
    \node[magenta] at (raw_int1.north) {(b) $0.03a$};
    \node[yshift=-1.2mm] at (exp_integrated_int.south) {(c)};
    \node[yshift=-1.2mm] at (dosdis_normalized.south) {(d)};
  \end{tikzpicture}
  
  \caption{\label{fig:dos}(a,b) Experimental normalized intensity
    spectra obtained by scanning along the waveguide position for two
    different amounts of extrinsic disorder as indicated in the
    figure. (c) Experimental spatially-integrated intensity for
    varying degrees of extrinsic disorder as labelled in the
    figure. (d) Calculated ensemble averaged DOS (normalized) for the
    fundamental waveguide mode for the six disordered samples in
    (c). The red-dashed line in the lowest disorder case is given by
    $A\frac{\lambda_{\rm min}^2}{\lambda^2}$ with $A=0.2$ and
    represents the qualitative contribution of radiation modes to the
    DOS. The amount of intrinsic disorder in all samples is
    $\sigma_{\rm i} = 0.005a(1.2\mathrm{nm})$.}
\end{figure}

Having identified the limits of perturbation theory above, we now
present some brute force calculations of the LDOS using full 3D FDTD
computations in a disordered PCW lattice. The numerical complexity is
very  demanding so we are restricted to much smaller waveguide
lengths than used in the experiment; also, we can only compute a small number of instances
which are not enough to compute the ensemble average trend shown in
Fig.~\ref{fig:dos}. This is mainly due to the large memory
requirements of the simulation volume since it cannot be reduced by
using symmetric boundary conditions due to symmetry breaking caused by
disorder.  Nevertheless, such calculations are useful for getting a
physical picture of what is happening for a particular instance and
section of a disordered PCW.

\begin{figure}
   \begin{tikzpicture}  
    \draw node (ldos_normalized) {\includegraphics[scale=0.55]{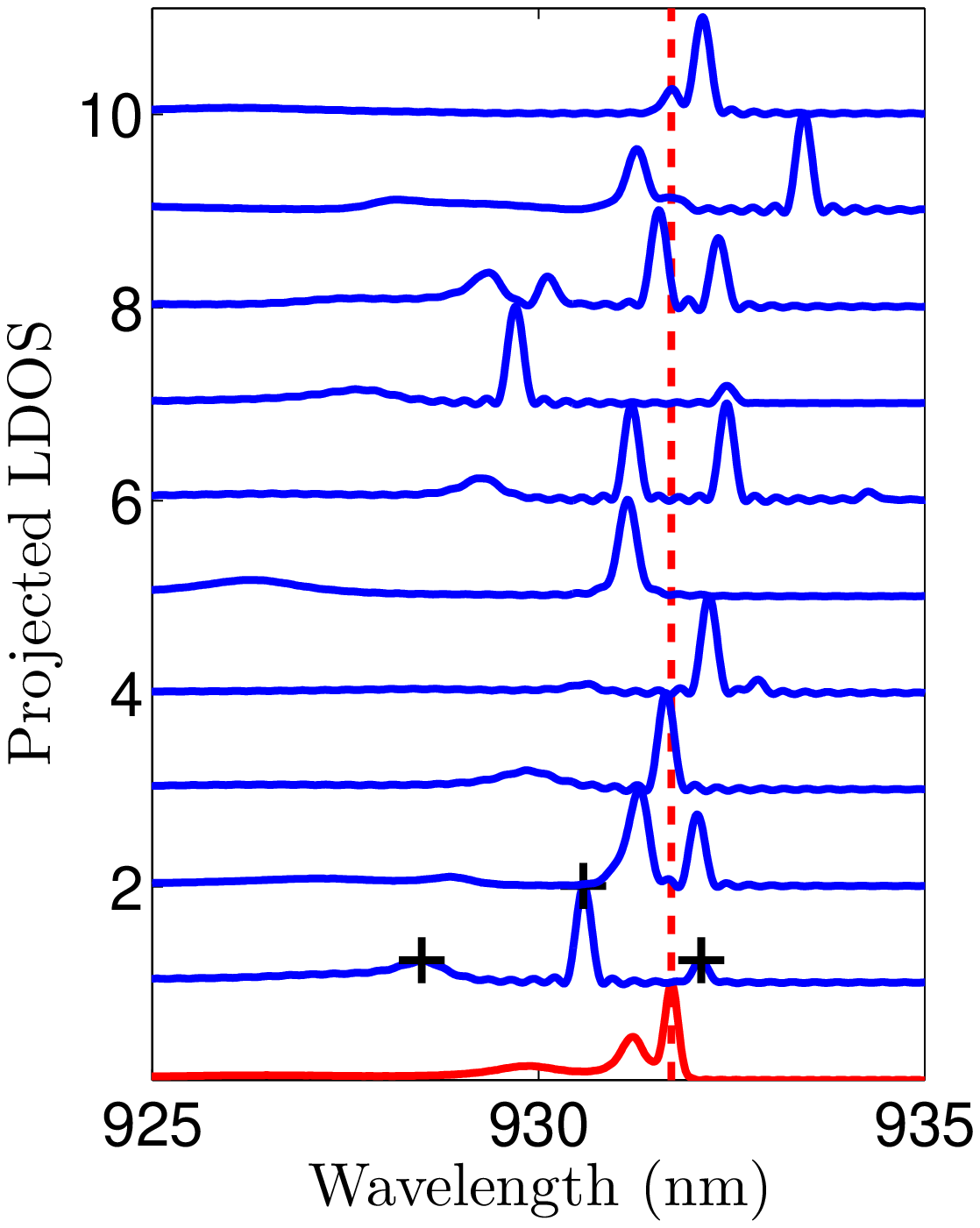}};
    \draw node [below=of ldos_normalized, yshift=12mm] (field1) {\hspace{-4.5mm}\includegraphics[scale=0.45]{{{fig_W1_distotal_0.005_0.02_E_ay2_N30_3D_1_33}}}}; 
    \draw node [below=of field1, yshift=12mm] (field2) {\includegraphics[scale=0.45]{{{fig_W1_distotal_0.005_0.02_E_ay2_N30_3D_1_22}}}};
    \draw node [below=of field2, yshift=12mm] (field3) {\hspace{-2.8mm}\includegraphics[scale=0.45]{{{fig_W1_distotal_0.005_0.02_E_ay2_N30_3D_1_15}}}};
  \end{tikzpicture}
  \caption{ (Colour online) \label{fig:ldos_instances_and_modes}(Top)
    Projected LDOS values of a $y$-oriented dipole centred at the
    anti-node of $e_y$, computed for ten instances of disorder
    (blue/dark-solid) with internal/external disorder values of
    $\sigma_{i}=0.005a(\SI{1.2}{\nano\metre}),
    \sigma_{e}=0.02a(\SI{4.8}{\nano\metre})$
    respectively. For reference, the LDOS with no \emph{extrinsic}
    disorder (red/light-solid) and ideal mode edge (dashed-red/light
    vertical line) are also shown. All LDOS instances are normalized
    to their own LDOS peak. The length of the waveguides was kept
    fixed at \SI{7.2}{\micro\metre} (30 unit cells). (Bottom) As
    highlighted by the black markers ($+$) on the Projected LDOS
    instance, starting from above (right) and going below (left) the mode-edge,
    localized mode intensity of $|e_y|^2$ is shown.}
      %(Top-Right) All LDOS instances (blue) are
      %normalized to the peak of the ideal LDOS (red) to give one an
      %idea of the expected Purcell factor enhancement. The ideal LDOS
      %peak corresponds to a Purcell factor of about $104.4$ and the
      %maximum Purcell factor observed for a disordered instance is
      %approximately $155.5$. 
      %The length of the waveguide is $30a(7.2)~\mu m$.}
\end{figure}

%To compute a finite length
%\SI{100}{\micro\metre} has $417$ unit cells is impractical in memory
%usage. The GF calculation requires a fine mesh and since we have
%Moreover, one must generate a lot of instances which is
%impractical.. What about Ergodic theory?
%--------------------Instances LDOS and modes-------------------------------%
To show that the DOS varies from instance to instance given the
disorder is kept fixed, we calculate the projected LDOS
$\rho_{\mathbf{\mu}}(\mathbf{r}, \omega)$ for ten statistically
disordered finite-length PCWs, as shown in
Fig.~\ref{fig:ldos_instances_and_modes}, by directly computing the
numerically exact photonic Green function of the PCW (see
Ref.~\onlinecite{Yao2009} for numerical implementation details) using
the 3D FDTD method \cite{lumerical}. The samples we simulate are only
$7.2$~$\mathrm{\mu}$m long ($30$ unit cells). With the waveguide cross
section in the $xy$ plane, denoting the waveguide direction as $x$ and
the origin at the centre of the waveguide, we compute the LDOS of a
$y$-oriented dipole $\rho_y(0, \omega)$ placed at the anti-node of
$e_y$, which occurs at the origin. The intrinsic/extrinsic disorder
values are
$ 0.005a(\SI{1.2}{\nano\metre}),\, 0.02a(\SI{4.8}{\nano\metre}),$
respectively. While ten instances which are only $30$ unit cells long
are not enough to conclude the existence of a mean blueshift of the
mode-edge, the variance in LDOS profiles can partially explain the
discrepancy observed in Fig.~\ref{fig:dos}(b) for
$\sigma_e=0.02a(\SI{4.8}{\nano\metre})$.  For completeness,
Fig.~\ref{fig:ldos_instances_and_modes} also shows examples of
disorder induced localized modes that appear both above and below the
mode edge. These modes are formed via multiple scattering in
cavity-like defects introduced via disorder.

We highlight that we have found a 2D FDTD method to be inadequate for
computing the Green function and LDOS for the PCW slab. Firstly, the
mode edge for a 2D PCW with the same structural parameters (apart from
the slab height) is different (\SI{1.2}{\micro\metre}) and secondly, a
2D PCW does not possess radiation or leaky modes and out-of-plane
decay cannot be computed. Hence the computed Green function does not
accurately capture the realistic 3D resonance shifts expected in the
LDOS. Although a 2D calculation can capture the qualitative modal
profile of the localized modes, their sensitivity to disorder is quite
different to 3D quasimodes. Thus one requires a 3D FDTD model to
compute the LDOS for a PCW slab in general.
%Lastly, there remains the question of the number of unit cells used
%and finite-size effects. The experimental waveguides are about
%$100$~$\mathrm{\mu}$m long (approximately $417$ unit cells) where the
%ones we simulate in Fig.~\ref{fig:ldos_instances} are only
%$7.2$~$\mathrm{\mu}$m long ($30$ unit cells). 
%We study the LDOS of an ideal
%W1 (no disorder) for two lengths as shown in Fig.~\ref{fig:ldos_lengths}. 
%As we increase the waveguide length, we see increased Fabry-Perot type
%reflections shown by the multiple peaks in the LDOS. Moreover, there are fast
%oscillations in the LDOS because our simulation time is short. Increasing
%simulation time will correct this but this approach is intractable as
%the number of unit cells increases. Therefore, to get an accurate LDOS
%for a large number of unit cells is intractable in 3D but our main point
%was to show the variance in mode-edge shifts which is done reasonably well
%with our current simulations.

%\begin{figure*}
  %\includegraphics[scale=0.3]{./figures/fig_ldos_y_relative_instance}\\
 % \includegraphics[scale=0.4]{{{./figures/fig_W1_distotal_0.005_0.02_E_ay2_N30_3D_1_33}}}
  %\includegraphics[scale=0.4]{{{./figures/fig_W1_distotal_0.005_0.02_E_ay2_N30_3D_1_22}}}
  %\includegraphics[scale=0.4]{{{./figures/fig_W1_distotal_0.005_0.02_E_ay2_N30_3D_1_15}}}

  %\caption{\label{fig:loc_modes} The localized modes for a
  %  disordered instance as highlighted by the black markers shown in
  %  the Fig.~\ref{fig:ldos_instances}.}
%\end{figure*}

\section{\label{sec:discussion}Discussion and Connections to Previous  Works}
%-----------------------------DOS-----------------------------------%

%-----------------Limits and strengths perturbation theory-----------%
%Our results agree qualitatively well with experimental data except for
%one case which is due to experiment being an instance as explained in
%previous section. 
%
%
%to compute the resonance shifts and the emergence of
%localized modes far from the mode-edge. 

Our theory, though perturbative, provides an intuitive and
computationally efficient semi-analytical approach to producing
experimentally relevant results for moderate amounts of extrinsic
disorder and provides upper bounds for high amount of extrinsic
disorder. One computation which includes computing the ideal Bloch
modes, Monte Carlo runs for the expectations and Riemann integrals
%and DOS sampling \com{how many wavelegths, what
 % graph?} 
for a given amount of disorder takes roughly $3$ hours on a
single-core CPU whereas computing the LDOS of a \SI{7.2}{\micro\metre}
long disordered PCW using 3D FDTD takes approximately $10$ CPU hours
for \emph{each} disorder instance on a cluster using $20$ multi-core
CPU nodes.
%Lastly ensemble averaged quantities are preferred than instance
%quantities when computing because they can be used further to produce
%other relevant quanities such as the the average PF enhancement in
%these disordered PCWs.

%------------Experiment sheds light on which pol model to use--------------%
As we have stated before, the bump-perturbation polarization model is
crucial to our findings. It is not the exact shape of the bump that is
important (see Refs. \cite{Johnson2005,Ramunno2009}) but the use of
asymmetric polarizabilities that yields a non-zero net mean frequency
shift. In the context of disorder-induced losses where all three
polarization models produce similar results, previously we have argued
that the bump polarization model should be best suited for modelling
disorder characterized via rapid radial fluctuations and the
smooth-perturbation model should be valid as long as the air-slab
interface remains nearly circular \cite{Patterson2010} which is indeed
the case considered in this work. To resolve this ambiguity, we rely
on the comparison with experimental findings (see
Sec.~\ref{sec:results}) which indicates that the bump-polarization
model (in the absence of any other models in the literature) is best
suited for all types of disorder and various disorder-induced
phenomena in PCWs.

%Given the experimental data and our findings, we conclude that
%the bump-polarization model should be used when modelling
%disorder-induced phenomena in photonic crystal waveguides.
%To solve this ambiguity, one must now rely on experimental data
%which shows us that the bump-polarization model should be used when
%modelling disorder-induced phenomena in photonic crystal waveguides.
%-----------------Other Literatute Works--------------------------%
Reference~\onlinecite{Patterson2010}  highlighted the importance of
accounting for local field effects in PCWs by computing
disorder-induced resonance shifts for the three polarization models
mentioned in Sec.~\ref{sec:shifts}. While the impact of disorder on
the DOS is qualitatively well known, to our knowledge, no one has
quantified the expected resonance shifts or spectral broadening as a
function of disorder and computed the disordered DOS which is found in
good qualitative agreement with experiments. Previously, spectral
broadening of the bandstructure was observed experimentally by Le
Thomas \emph{et al.} \cite{LeThomas2009} and predicted theoretically
by Savona \cite{Savona2011} whose findings showed increased spectral
broadening of the bandstructure as the disorder increases, but did not
predict a blueshift or quantify the expected shift or broadening as a
function of disorder.

%on inherently Savona computed the ensemble
%averaged disordered DOS by directly computing the disordered modes for
%multiple instances of a disordered PCW \cite{Savona2011}.
%But since the amount of disorder is not varied, one cannot
%conclude the existence of a pronounced blueshift.
%, on the contrary there seems
%a slight redshift in the resonances. READ LE THOMAS PAPER!

%--------------Importance of the model--------------------------%
\section{\label{sec:conclusions}Conclusions}
Through theory and experiment, we have shown that accurate modelling
of local field effects is critical for computing experimentally
relevant mean frequency shifts and realistic DOS profiles in
PCWs. These findings also point out the possible limitations of
disorder polarization models that do not include local field effects
or include local field effects through the use of symmetric
polarizabilities. For moderate amounts of disorder, our
computationally efficient semi-analytical perturbative approach yields
results that are in good qualitative agreement with experiments and
can be used to compute realistic quantities such as Purcell factor
enhancements in PCWs with embedded quantum dots.  We have also shown
examples of the numerically exact LDOS for various disordered
instances and the underlying disordered-induced resonance modes on a
small PCW using a rigorous 3D FDTD approach. Future work will focus on
developing a non-perturbative approach that takes into account
multiple scattering to better model high amounts of disorder.

\acknowledgements{This work was supported by the National Science and
  Engineering Research Council of Canada and Queen's University, Canada. Alisa Javadi,
  P.D. Garc\'{i}a and Peter Lodahl gratefully acknowledge the
  financial support from the Danish Council For Independent Research
  (Natural Sciences and Technology and Production Sciences), The
  Villum Foundation, and the European Research Council (ERC
  consolidator Grant ``ALLQUANTUM''). }

\bibliographystyle{apsrev4-1}
%\vspace{0.3cm}
\bibliography{references}
\end{document}